\documentclass[final,5p,times,twocolumn]{elsarticle}

\usepackage{graphicx}
\usepackage{amssymb}

\journal{Physics Letters B}

\newcommand{\oeq}{\begin{equation}}
\newcommand{\ceq}{\end{equation}}
\newcommand{\oeqn}{\begin{eqnarray}}
\newcommand{\ceqn}{\end{eqnarray}}

\renewcommand{\>}{\rangle}

\newcommand{\hb}{\hbar}

\begin{document}

\begin{frontmatter}

\title{Influence of entrance-channel magicity and isospin on quasi-fission}

\author[ANU,CEA]{C. Simenel\corref{cor1}}
\ead{cedric.simenel@cea.fr}
\cortext[cor1]{corresponding author}
\author[ANU]{D. J. Hinde}
\author[ANU]{R. du Rietz}
\author[ANU]{M. Dasgupta}
\author[ANU]{M. Evers}
\author[ANU]{C. J. Lin}
\author[ANU]{D. H. Luong}
\author[ANU]{A. Wakhle}
\address[ANU]{Department of Nuclear Physics, Research School of Physics and Engineering, Australian National University, Canberra, ACT 0200, Australia}
\address[CEA]{CEA, Centre de Saclay, IRFU/Service de Physique Nucl\'eaire, F-91191 Gif-sur-Yvette, France.}

\begin{abstract}
The role of spherical quantum shells in the competition between fusion and quasi-fission is studied for reactions forming heavy elements.
Measurements of fission fragment mass distributions for different reactions leading to similar compound nuclei
have been made near the fusion barrier.
In general, more quasi-fission is observed for reactions with non-magic nuclei.
However, the $^{40}$Ca+$^{208}$Pb reaction is an exception, showing strong evidence for quasi-fission, though both nuclei are doubly magic. Time-dependent Hartree-Fock calculations predict fast equilibration of $N/Z$ in the two fragments early in the collision.
This transfer of nucleons breaks the shell effect, causing this reaction to behave more like a non-magic one in the competition between fusion and quasi-fission.
Future measurements of fission in reactions with exotic beams should be able to test this idea with larger $N/Z$ asymmetries.
\end{abstract}

\begin{keyword}
Nuclear fusion \sep Quasi-fission \sep Heavy-element formation \sep Time-Dependent Hartree-Fock theory
\end{keyword}

\end{frontmatter}

Quantum shell effects play a key role in the structure and stability of atomic nuclei, as they do in the periodic chemical properties of the elements.
Where there is a large energy gap to the next quantum level, the total number of protons or neutrons filling all levels below the gap is referred to as a magic number.
In particular, magic nuclei have a smaller mass per nucleon than their neighbours.
The variation of the magic numbers across the nuclear chart is crucial to build our understanding of the nuclear quantum many-body system.
One major challenge is to define the magic numbers
in the region of the superheavy elements (SHE), with $Z\ge110$ protons~\cite{hof00,mor07,oga06,hof07}.
Associated with this, atom-by-atom measurements of the chemical properties of SHE are testing the predicted strong relativistic
effects on the  electrons  which modify the periodic Table~\cite{eic07}.

SHE up to $Z=118$ have been synthesised in fusion reactions of heavy nuclei, either using $^{208}$Pb and $^{209}$Bi targets~\cite{hof00,mor07}, or $^{48}$Ca beams on actinide targets~\cite{oga06,hof07}.
Production cross sections are, however, extremely small (of the order of a few picobarns), and a good understanding of the reaction mechanisms is needed to optimise their production.
To achieve a comprehensive global picture of SHE formation is very challenging, as many variables
may affect
fusion probabilities. These include collision energy, mass-asymmetry, deformation and orientation, isospin, and shell structure of the colliding nuclei.
These variables are often strongly entangled, making it difficult to isolate the effect of a single variable. Furthermore,
these properties evolve dynamically, thus
it is necessary to understand the different associated time scales.

\begin{figure*}
\begin{center}
\includegraphics[width=18cm]{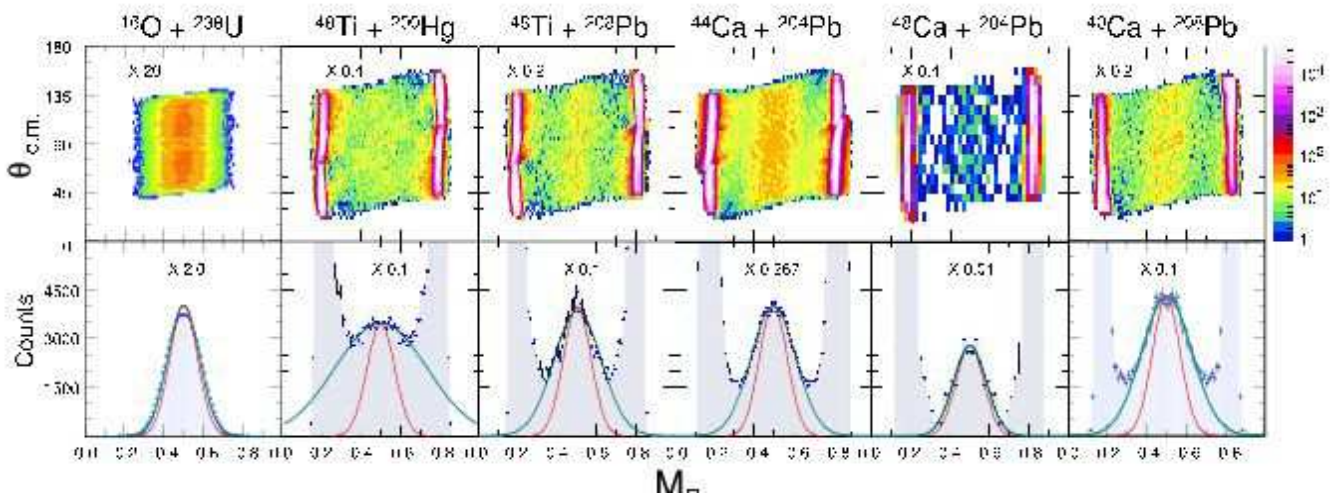}
\caption{(colour online). Measured mass-angle distributions for each reaction (upper panels). The factor multiplies the maximum counts of the logarithmic colour scale (right). In the projected mass ratio spectra
(lower panels) the scale factor multiplies the counts scale on the left. The difference between the scale factors is due to the various statistics obtained with each system. Gaussian fits to the region around $M_R$=0.5 are shown (turquoise lines), whose standard deviations $\sigma$ are given in Table~\ref{tab:width}. Gaussian functions with $\sigma_{MR}=0.07$ (thin red lines) are shown for reference.}
\label{fig:MAD}
\end{center}
\end{figure*}

The early stage of the collision is a crucial step in SHE formation, where the initial conditions are the most important. These determine the configuration
at which the colliding partners have dissipated their kinetic energy, thus determining the shape of the initial fragile dinucleus.  This can break apart,
generally after multiple nucleon transfers (mainly from the heavy to the light partner), in a process called quasi-fission (QF)~\cite{boc82,tok85,she87}.
Alternatively it may reach compact shapes, fusing to form a hot compound nucleus (CN), which can lead to formation of a SHE through neutron evaporation in competition with CN fission.
Although the CN survival probability against fission is very small, its decay width is governed by the well-known equations for statistical decay, which should allow prediction of the
relative survival probabilities from different fusion reactions.
This is not the case for QF, which is a completely dynamical process, and depends on many variables which can be different for different reactions.
The nature of the two fission processes are reflected in their time scales,
which can be very different. Typical times scales for QF are shorter than $10^{-20}$~s~\cite{boc82,tok85,she87,rie11},
but can be longer than $10^{-16}$~s~\cite{and07} for fusion-fission.
To efficiently form SHE, the entrance channel conditions should be chosen to minimise the QF probability, which is dominant in reactions forming SHE.
Beyond the basic principle of minimising the Coulomb energy in the entrance channel, a quantitative understanding of the effects of nuclear structure on
the competition between fusion and QF is a key missing ingredient.

At collision energies above the fusion barrier,
a systematic analysis showed that
closed shells in the
colliding nuclei
have a relatively small effect on fusion probabilities~\cite{hin05}.
However, at energies around the barrier,
the competition between fusion and QF is known to be affected by (shell-driven) nuclear deformation and orientation~\cite{hin95,liu95,hin96,oga04,kny07,hin08,nis08}.
Spherical shells may also be important, resulting in so-called ``cold valleys'' in the potential energy surface, which lead to the compact CN configuration~\cite{san76,gup77,faz05,ari06}.
Fusion through these valleys may also be favoured because energy dissipation should be weaker, allowing greater inter-penetration before the initial kinetic energy is
dissipated~\cite{hin05,arm00}.
These effects may be vital in the recent successful synthesis of SHE~\cite{hof00,mor07,oga06,hof07}.
However, the interplay of spherical shells with other degrees of freedom, such as the isospin of the two colliding nuclei, has not yet been investigated.

In this letter, the role of spherical shells (magicity) on the QF probability is first demonstrated through fission
measurements for reactions with relatively small initial isospin asymmetry, or more precisely $N/Z$ asymmetry,
quantified by the difference between the $N/Z$ ratios of the initial colliding nuclei $\Delta{(N/Z)}_i$.
Then, the case of a magic reaction with large $\Delta{(N/Z)}_i$ is investigated.
The time scales for QF and isospin equilibration are investigated with the help of calculations,
and used to explain the measurements in terms of the dynamical interplay between isospin asymmetry and spherical shells.

Measurements were made using the 14UD electrostatic accelerator at the Australian National University.
Pulsed beams of 111~MeV $^{16}$O and 213.5~MeV $^{40}$Ca, and DC beams (giving higher intensities) of 212~MeV $^{44}$Ca, 213~MeV $^{48}$Ca,
and 230, 235 MeV $^{48}$Ti were produced from metallic $^{nat}$Ca and $^{nat}$Ti samples.
Isotopically enriched targets of $^{204}$Pb (420~$\mu$g/cm$^2$ self-supporting), and $^{208}$PbS (30~$\mu$g/cm$^2$),
$^{200}$Hg (15~$\mu$g/cm$^2$) and $^{238}$UF$_{4}$ (400~$\mu$g/cm$^2$), evaporated onto $\sim$15~$\mu$g/cm$^2$ $^{nat}$C backings, were mounted
on a target ladder whose normal was at 60$\,^{\circ}$ to the beam.
Binary reaction products were detected in coincidence using two 28$\times$36~cm$^2$ position-sensitive multiwire proportional
counters on opposite sides of the beam, covering laboratory scattering angles of
$5\,^{\circ}<\theta<80\,^{\circ}$ and $50\,^{\circ}<\theta<125\,^{\circ}$.
For the pulsed beams, the measured positions and times-of-flight allowed direct reconstruction of the fragment velocities~\cite{hin96}.
With DC beams, the velocities were determined from the time difference between the two fragments~\cite{tho08}, assuming binary reactions and full momentum transfer,
which will be valid for the low fissility targets used~\cite{hin96}.
Following iterative correction for energy loss in the target, the mass ratio $M_R=m_1/(m_1+m_2)$
(where $m_1$ and $m_2$ are the two fragment masses)
and the centre-of-mass (c.m.) scattering angle $\theta_{c.m.}$ were deduced.
Since both fragments are detected, the mass-angle distribution (MAD) is populated twice~\cite{tho08}, at ($M_R, \theta_{c.m.})$ and $(1-M_R, \pi-\theta_{c.m.}$).

The MAD for the reactions measured are shown in the upper panels of Fig.~\ref{fig:MAD}. 
The reactions with Ca and Ti beams form isotopes of the elements No (Z=102) and Rf (Z=104), and involve 
similar charge products in the entrance-channel. The $^{16}$O+$^{238}$U reaction forms Fm (Z=100),
but with less than half the entrance-channel charge product.
In the measurement the azimuthal coincidence coverage was essentially 90$\,^{\circ}$ for all~$\theta$,
thus the number of events in each MAD bin is proportional to the angular differential cross section d$\sigma$/d$\theta_{c.m.}$.
Note, however, that every MAD has a different coefficient of proportionality due to the varying statistics obtained for each reaction.
The intense bands at extreme $M_R$ values correspond to elastic and quasi-elastic (QE) scattering, while
fission-like events, associated with either
fusion-fission or QF, are spread around $M_R=0.5$.
Note that, in our measurements with Ca and Ti beams, both fusion and QF occur at similar partial waves. 
Indeed, the beam energies correspond to below-barrier energies, as can be seen from Table~Ê\ref{tab:width}, where centre-of-mass energies and calculated barrier energies are given. Thus the angular momenta involved are low, and those of fusion and quasi-fission are bound to show a large overlap.

For the heavier projectiles, the fission-like events clearly show a correlation of fragment mass with angle,
resulting from the short reaction times ($\le10^{-20}$~s)~\cite{tok85,she87,rie11}.
For example, for the $^{44}$Ca+$^{204}$Pb reaction, the $M_R$ centroid for $125\,^\circ<\theta_{c.m.}<135\,^\circ$ is $0.511\pm0.004$.
Although the deviation from $M_R$=0.5 is small, it is much larger than the statistical uncertainty,
and as clearly seen in the MAD, varies consistently with $\theta_{c.m.}$. Reference measurements for the 
reactions of $^{16}$O with $^{208}$Pb, and with $^{238}$U (shown in the left-most panel of Fig.~\ref{fig:MAD}) give 
essentially no correlation of mass with angle, consistent with much longer fission times.

The lower panels of Fig.~\ref{fig:MAD} show the $M_R$ projections of the MAD spectra above. 
The widths of the fission-like fragment mass distributions are expected to be larger in the presence of QF than if only fusion-fission is present~\cite{she87,hin08,bac96}.
To characterise the $M_R$ distributions for the fission-like events, and to allow comparison with previous work~\cite{pro08}, they were fitted with Gaussian functions, 
within the range 0.34$\leq$$M_R$$\leq$0.66 (turquoise curves in Fig.~\ref{fig:MAD}) so as to exclude deep-inelastic and QE events.
For $^{16}$O+$^{238}$U, we choose 0.2$\leq$$M_R$$\leq$0.8 as only fission-like events were detected.
The fitted standard deviations $\sigma_{MR}$ are given in Table~\ref{tab:width},
together with the value for 218 MeV $^{48}$Ca+$^{208}$Pb from Ref.~\cite{pro08}.
Since it may well be that the true distributions are not single Gaussians~\cite{hin96,hin08,tho08,pro08,itk04,itk11},
we also compute the standard deviation $\Sigma_{MR}$ of the data points in the same 0.34$\leq$$M_R$$\leq$0.66 range, 
which are also given in Table~\ref{tab:width}.
Of course, $\sigma_{MR}$ and $\Sigma_{MR}$ are different quantities with different values, but 
they both constitute a measure of the width of the fission-like fragment mass distributions, the latter independent of any assumed shape. 
As will be seen, the two quantities do exhibit the same trends, and together with the reasonable reproduction of the experimental data by the Gaussian fits,
suggest that the fitted standard deviations $\sigma_{MR}$ give a reasonable representation of the mass width of the fission-like events, with a single parameter.

\begin{table*}
\begin{center}
\caption{\label{tab:width}Standard deviation $\Sigma_{MR}$ of fission-like fragment mass distributions and standard deviation $\sigma_{MR}$ of their Gaussian fits (see text) for each reaction, with statistical uncertainties. $N_m$ is the total number of magic numbers in target and projectile, and $\Delta{(N/Z)}_i$ is the difference between their $\frac{N}{Z}$ ratios. Centre-of-mass energies $E_{c.m.}$ and theoretical barriers $B_{th}$ from the proximity model~\cite{swi05} are in MeV.}
\begin{tabular}{cccccccccc}
reaction                        & CN      &  $E_{c.m.}$& $B_{th}$  &$N_m$ &$\Delta{(\frac{N}{Z})}_i$   & $\sigma_{MR}$ & $\Sigma_{MR}$ & \\
\hline
$^{16}$O+$^{238}$U & $^{254}$Fm  & 104.0  & 80.3 & 2         &  0.59    &   $0.081\pm0.001$ & $0.073\pm0.001$ & Present work\\
$^{48}$Ti+$^{200}$Hg & $^{248}$No  & 185.5  & 190.9 & 0         &  0.32    &   $0.237\pm0.025$ & $0.090\pm0.001$& Present work\\
$^{48}$Ti+$^{208}$Pb  & $^{256}$Rf & 190.9   & 194.4 &2         &  0.35         &   $0.121\pm0.004$ & $0.082\pm0.001$& Present work\\
$^{44}$Ca+$^{204}$Pb & $^{248}$No & 174.4  & 178.0 &2        &   0.29     &   $0.114\pm0.002$ & $0.081\pm 0.001$& Present work\\
$^{48}$Ca+$^{204}$Pb & $^{252}$No &  172.4 & 175.8 &3        &  0.09   &   $0.084\pm0.008$ & $0.073 \pm 0.004$& Present work\\
$^{40}$Ca+$^{208}$Pb &  $^{248}$No &179.1  & 179.5 &4        &  0.54      &   $0.126\pm0.004$ & $0.083\pm0.001$ & Present work\\
$^{48}$Ca+$^{208}$Pb & $^{256}$No & 177.1  & 175.0 &4        &   0.14        &   $0.068\pm0.002$ & $0.064\pm 0.002$& From Ref.~\cite{pro08}\\
\end{tabular}
\end{center}
\end{table*}

\begin{figure}
\begin{center}
\includegraphics[width=7cm]{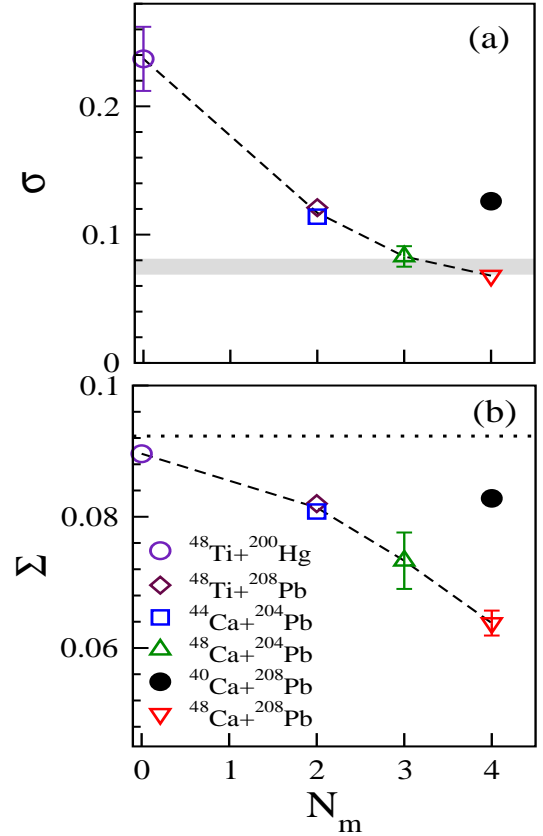}
\caption{(colour online). (a) Standard deviations $\sigma_{MR}$ of the Gaussian fits of the fission-like fragment mass distributions
as a function of
the number of magic numbers in the entrance channel
$N_m$.
The horizontal band shows the upper limit of $\sigma_{MR}$ for pure fusion-fission (i.e., without QF).
(b) Standard deviation $\Sigma_{MR}$
in the 0.34$\leq$$M_R$$\leq$0.66 range. The dotted line corresponds to a flat distribution.
When not shown, the statistical uncertainties are smaller than the size of the points.
The dashed lines guide the eye. }
\label{fig:sigma}
\end{center}
\end{figure}

In order to investigate the influence on quasi-fission of spherical shells in the entrance channel, we plot in
Fig.~\ref{fig:sigma} the widths ($\sigma_{MR}$ and $\Sigma_{MR}$) of the fission-like fragment distributions  as a function of the number $N_m$ of entrance channel magic numbers (given in Table~\ref{tab:width}).
The possible proton and neutron magic numbers for projectile and target nuclei are $Z=20$ and $N=20,28$, and $Z=82$ and $N=126$ respectively.
An upper limit to the standard deviation $\sigma_{MR}$ for fusion-fission ($\sigma_{fiss}$) of 0.07-0.08 can be taken from the present and previous~\cite{she87,hin96} measurements for $^{16}$O+$^{238}$U. This is only an upper limit as it was shown~\cite{hin96} that QF contributes to fission-like events even in this reaction. This
range is indicated by the horizontal band in Fig.~\ref{fig:sigma}(a).
Only the $^{48}$Ca+$^{204,208}$Pb data lie in this range. All other reactions have larger widths, indicating the presence of QF~\cite{she87,hin08,bac96}, 
a result consistent with the observation of a dependence of mean fragment mass with angle in the measured MAD presented in Fig.~\ref{fig:MAD}.

Apart from the $^{40}$Ca+$^{208}$Pb reaction, discussed later, a clear correlation is seen in Fig.~\ref{fig:sigma} between the entrance channel magicity, quantified by $N_m$, and the amount of QF, related to $\sigma_{MR}-\sigma_{fiss}$:
the less entrance-channel magicity, the more QF. As discussed in the introduction, 
this correlation could result from cold valleys in the potential energy surface ~\cite{san76,gup77,faz05,ari06} and/or weaker energy dissipation~\cite{hin05,arm00},
both effects being associated with the spherical shells.
As a result, a greater inter-penetration of the two nuclei should then be achieved,
leading to a higher fusion probability, and, consequently, a smaller QF probability.
This interpretation is also supported by the observation of relatively high fusion-evaporation cross-sections (up to $\sim3$~$\mu{b}$) in the $^{48}$Ca+$^{208}$Pb system as compared to reactions with non-magic targets with similar masses~\cite{gag89}.

Before accepting this conclusion, the possible effects of a number of additional variables should be considered. The comparison of the widths is strictly valid for reactions forming the same CN (here, the three reactions forming $^{248}$No), under the same conditions. The known dependence of the standard deviation $\sigma_{MR}$ on excitation energy for these reactions is too weak~\cite{hin96,pro08}, and the difference in energies too small (e.g., excitation energies in $^{48}$Ti+$^{200}$Hg and $^{44}$Ca+$^{204}$Pb differ by only 0.4 MeV) for differences in excitation energy to affect the conclusions. The $^{48}$Ti+$^{208}$Pb reaction has the largest entrance-channel charge product, and forms the heaviest and most fissile nucleus, thus without shell effects, the largest standard deviation $\sigma_{MR}$ might be expected. This is not what is observed, so we conclude that the large changes in $\sigma_{MR}$ must be related to the differing magicity in the entrance channel, rather than properties of the composite system.

There is one reaction that does not follow the systematic behaviour shown by the others, namely $^{40}$Ca+$^{208}$Pb.
Fig.~\ref{fig:sigma} shows clearly that it
demonstrates strong evidence for QF ($\sigma_{MR}\simeq0.13$), despite having maximal magicity $N_m=4$.
We propose an explanation below which does not invalidate the link between magicity and QF probability seen for the other reactions.
To solve this puzzle, it is sufficient to invoke the fast isospin equilibration resulting from nucleon transfer.
Detailed measurements of reaction product yields~\cite{kro10} and angular distributions~\cite{szi05} have shown that systems with strong isospin asymmetry in the entrance channel (like $^{40}$Ca+$^{208}$Pb~~\cite{szi05})
undergo a rapid (although incomplete) isospin equilibration in the early stage of the collision, through the transfer of nucleons~\cite{szi05}.

The  time-dependent Hartree-Fock (TDHF) theory has successfully described transfer in $N/Z$ asymmetric reactions (for example Refs.~\cite{bon81,sim01,sim07,iwa10}).
Here, it is used to investigate the timescale of isospin equilibration via transfer.
In  TDHF, each particle evolves independently in the mean-field generated by all the others.
The TDHF formalism is optimised for the prediction of expectation values of one-body operators, such as the average $N/Z$ ratio in the fragments.
The \textsc{tdhf3d} code is used with the SLy4$d$ parameterisation of the Skyrme functional~\cite{kim97}.
The TDHF equation is solved iteratively in time, with a time step $\Delta{t}=1.5\times10^{-24}$~s, on a spatial grid of $56\times56\times28/2$ points with a plane of symmetry (the collision plane), and a mesh size $\Delta{x}=0.8$~fm
(see~\cite{sim10a} for more details).
The initial distance between the nuclei is 22.4~fm.

\begin{figure}
\includegraphics[width=8cm]{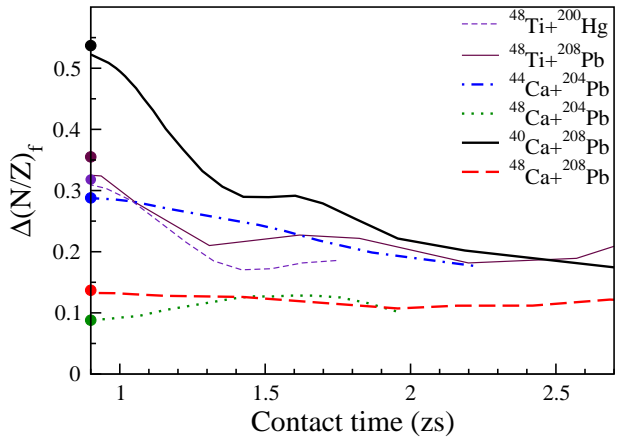}
\caption{(colour online). Final $N/Z$ asymmetry
of the fragments (lines) as a function of contact time of the fragments in zeptoseconds (1~zs=10$^{-21}$~s), defined as the time during which the neck density exceeds half the saturation density $\rho_0/2=0.08$~fm$^{-3}$. The initial values $\Delta(N/Z)_i$ are shown by full circles.}
\label{fig:NZ}
\end{figure}

The results of the TDHF calculations of $N/Z$ equilibration between two colliding nuclei are shown in Fig.~\ref{fig:NZ}, as a function of their contact time.
The difference in the $N/Z$ ratios of the two nuclei before any transfer of nucleons is denoted by $\Delta{(N/Z)}_i$, shown by the full circles in Fig.~\ref{fig:NZ}, and also given in Table~\ref{tab:width}. The curves show the calculated evolution of the difference between the $N/Z$ ratios of the outgoing (final) fragments ($\Delta{(N/Z)}_f$) for each reaction, as a function of contact time, defined as the time during which the neck density exceeds half the saturation density $\rho_0/2=0.08$~fm$^{-3}$.
The contact time is varied by making calculations at angular momentum~$L\hb$ from $\sim20\hb$ to 70$\hb$. The energies of the collisions are the same as in the experiment. For $L<20$, most of the systems undergo capture resulting in fusion, whose timescales are too long for the TDHF calculations, or strongly damped collisions. For the smallest contact times (associated with large $L$), the nuclei scatter (in)elastically and no change in isospin occurs (i.e., $\Delta{(N/Z)}_f\simeq\Delta{(N/Z)}_i$) as seen in Fig.~\ref{fig:NZ}. 
For the $^{48}$Ca+$^{204,208}$Pb reactions, the initial isospin asymmetry is small, and no change in isospin occurs with increasing contact times. 
The fact that, for these reactions, $\Delta{(N/Z)}_f$ never reaches zero is typical for mass asymmetric reactions~\cite{pla90}.
For the other reactions, as the contact time increases, the $\Delta{(N/Z)}_f$ approaches the same isospin asymmetry.
In particular the most $N/Z$ asymmetric reaction, $^{40}$Ca+$^{208}$Pb
undergoes a large reduction of $\Delta{(N/Z)}_f$, in agreement with experiment~\cite{szi05}.
Using a particle number projection technique~\cite{sim10b}, the most probable outcome for this reaction after a contact time of $\sim2.7\times10^{-21}$~s (calculated for $L=20$), is found to be $^{42}$Ar+$^{206}$Po. This calculation also gives the probability of remaining in the entrance channel (and thus of conserving its entrance channel magicity), which is $P_{\Delta{Z}=0}P_{\Delta{N}=0}\simeq0.083\times0.002\simeq1.7\times 10^{-4}$, a negligible probability.
However, for the $^{48}$Ca+$^{208}$Pb reaction, even for a contact time as long as $\sim3.5\times10^{-21}$~s (not shown in Fig~\ref{fig:NZ}), this probability is still $0.76\times0.57\simeq0.43$, giving a much larger survival probability for the initial magic numbers in this reaction.

\begin{figure}
\includegraphics[width=8cm]{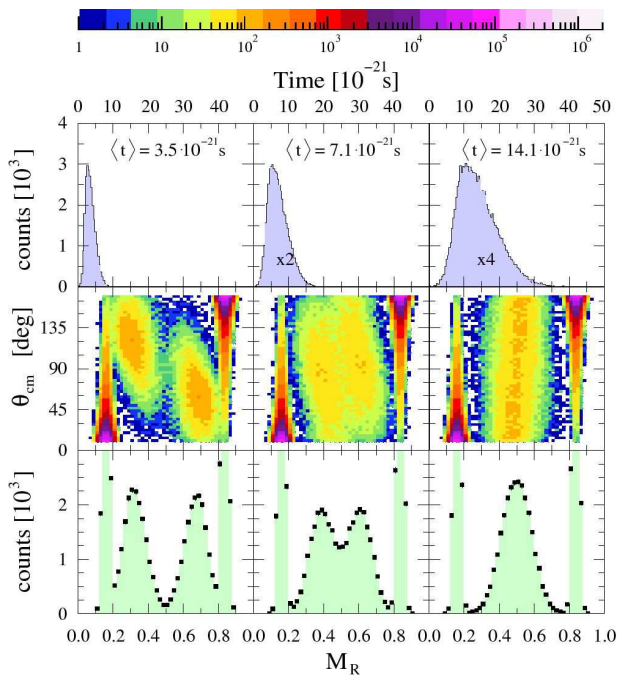}
\caption{(colour online). Simulated mass-angle distributions (middle panels) for  $^{40}$Ca+$^{208}$Pb  following~\cite{rie11}, for the quasi-fission time distributions shown in the upper panels. The average time $\<t\>$ of these distributions is indicated. The projections on the mass ratio axis are shown in the lower panels.}
\label{fig:sim}
\end{figure}

If the nucleons are transferred and the magicity is lost early in the collision, the system should behave more like a non magic system. On the contrary, if isospin equilibration takes place on a time scale similar to that of QF, then the magicity in the entrance channel could still significantly enhance fusion. According to the TDHF calculations (Fig.~\ref{fig:NZ}),  $^{40}$Ca+$^{208}$Pb   experiences a high degree of isospin equilibration for sticking times $\le2\times10^{-21}$~s. This is in agreement with experimental observations~\cite{szi05} of a high degree of N/Z equilibration in deep-inelastic collisions before many nucleons have been exchanged. This time has to be compared with the typical time scale for QF. Microscopic quantum theories cannot yet model such collisions from first principles~\cite{sim11}. Thus, to obtain the QF time for the reactions studied, MAD have been simulated using a classical trajectory model~\cite{rie11}. MAD were calculated for three different QF time distributions shown in the upper panels in Fig.~\ref{fig:sim}, whose mean times varied from 3.5$\times10^{-21}$~s to 14$\times10^{-21}$~s. The calculated MADs corresponding to these mean times are shown in the middle panels of Fig.~\ref{fig:sim}, whilst the bottom panels show the predicted mass ratio spectra. The shape of the experimental data (Fig.~\ref{fig:MAD} right panel) is best reproduced with an average time scale of 14$\times10^{-21}$~s, which is much longer than the time for
isospin equilibration. Isospin equilibration leading to loss of magicity occurs early in the $^{40}$Ca+$^{208}$Pb collision, which thus may be expected to exhibit QF properties closer to non magic systems. This is what is seen experimentally, as clearly shown in Fig.~\ref{fig:sigma}.

Finally, let us note that previous measurements~\cite{boc82,pro08,pac92} of excitation functions for capture reactions (including both fusion-fission and quasi-fission processes) in $^{40,48}$Ca+$^{208}$Pb have shown different behaviours in the two systems. 
In particular, at sub-barrier energies, reactions induced by $^{40}$Ca were found to produce larger capture cross-sections~\cite{pac92}.
This increase is consistent with our interpretation which is that this is a result of positive Q-value transfer reactions associated with 
$N/Z$ equilibration in the $^{40}$Ca reactions.

To conclude, experimental MAD for reactions with small isospin asymmetry show that magic numbers in the entrance channel reduce quasi-fission and are thus expected to increase the probability for fusion, while non magic systems show more quasi-fission. With a large initial isospin asymmetry, a rapid $N/Z$ equilibration occurs in the early stage of the reaction, modifying the identities of the collision partners.
This is the case for  $^{40}$Ca+$^{208}$Pb, which, as far as the competition between fusion and quasi-fission is concerned, behaves more like a non magic system, i.e., with increased quasi-fission.
Reactions with the neutron-rich $^{48}$Ca on heavy targets usually have small isospin asymmetry, and thus are more favourable to fusion than reactions with $^{40}$Ca, as
well as leading to more neutron-rich compound nuclei having a higher probability of surviving fusion-fission.
The importance of isospin asymmetry in the entrance channel should be considered in planning fusion experiments with exotic beams to form and study new 
 isotopes of existing elements, as well as new super-heavy elements. 

The authors are grateful to N. Rowley for a careful reading of the manuscript and for insightful comments. N. Lobanov and D. C. Weisser are thanked for intensive ion source development.
The TDHF calculations were performed on the NCI National Facility
in Canberra, supported by the Commonwealth Government.
Support from ARC Discovery grants DP06644077 and DP110102858 is acknowledged.

\end{document}